\documentclass[12pt]{article}
\usepackage{graphicx}
\usepackage{natbib}
\begin{document}

\begin{center}
{\large {\bf
Study of young stellar groupings in H\,{\sc ii}~regions based on the 
spectral and photometric data
}}

\bigskip

A.S.~Gusev$^{1}$\footnote{Email: gusev@sai.msu.ru}, 
F.Kh.~Sakhibov$^{2}$, A.E.~Piskunov$^{3}$, N.V.~Kharchenko$^{4}$, 
L.S.~Pilyugin$^{4}$, O.V.~Ezhkova$^{1}$, M.S.~Khramtsova$^{3}$, 
S.A.~Guslyakova$^{5}$, V.V.~Bruevich$^{1}$, S.N.~Dodonov$^{6}$, 
V.~Lang$^{2}$, E.V.~Shimanovskaya$^{1}$, Yu.N.~Efremov$^{1}$

\bigskip

$^1${\it Sternberg Astronomical Institute, Moscow State University, 
Moscow, Russia}

$^2${\it University of Applied Sciences of Mittelhessen, Department of 
Mathematics, Natural Sciences and Data Processing, Friedberg, 
Germany}

$^3${\it Institute of Astronomy, Russian Academy of Sciences, 
Moscow, Russia}

$^4${\it Main Astronomical Observatory of National Academy of Sciences 
of Ukraine, Kiev, Ukraine}

$^5${\it Space Research Institute, Russian Academy of Sciences, 
Moscow, Russia}

$^6${\it Special Astrophysical Observatory, Russian Academy of Sciences, 
Nizhnij Arkhyz, Russia}

\bigskip

{\small {\it
(Received 15 September 2015)
}}

\bigskip

\end{center}

{\small
We studied 102 star forming regions in seven spiral galaxies (NGC~628, 
NGC~783, NGC~2336, NGC~6217, NGC~6946, NGC~7331, and NGC~7678) on the 
basis of complex spectroscopic, photometric ($UBVRI$) and 
spectrophotometric (H$\alpha$ line) observations. Using data on the 
chemical composition and absorption in H\,{\sc ii}~regions, obtained from 
spectroscopic observations, and using evolutionary models, we estimated 
physical parameters (ages and masses) of young stellar groupings embedded 
in H\,{\sc ii}~regions. We found that the gas extinction, $A{\rm (gas)}$, 
which determined from the Balmer decrement, does not correspond in some 
cases to the absorption $A{\rm (stars)}$ in the young stellar associations 
(complexes). This is due to the spatial offset relative H\,{\sc ii}~cloud 
the stellar group related to him. It has been found that the condition 
$A{\rm (gas)} = A{\rm (stars)}$ does not satisfied for the star 
forming regions, in which: 1) the contribution to the total emission of 
gas in the $B$ and/or $V$ bands is higher than $40\%$, and 2) 
EW(H$\alpha)>1500$\AA. Extinction $A(V)$ in studied star forming regions 
corrected for the Galactic absorption $A(V)_{\rm Gal}$ ranges from 0 to 
3~mag with a mean value $A(V) - A(V)_{\rm Gal} = 1.18\pm0.84$~mag. We 
estimated masses and ages for 63 star forming regions. The regions have ages 
from 1 to 10~Myr, the most part of them are younger than 6~Myr. The derived 
masses of young stellar groupings range from $10^4 M_{\odot}$ in the nearby 
galaxies NGC~628 and NGC~6946 to $10^7 M_{\odot}$ in the most distant 
NGC~7678. More than $80\%$ of groupings have masses between $10^5 M_{\odot}$ 
and $10^6 M_{\odot}$. The lowest mass estimate of 
$\approx1\cdot10^4 M_{\odot}$ for the objects in NGC~628 and NGC~6946 
belongs to the mass interval of the youngest Galactic open clusters.

\bigskip

}

\section{Introduction}

We present the results of comprehensive study of star forming (SF) regions 
in seven spiral galaxies based on the obtained spectral and photometric 
data. We have observed emission line spectra of 102 giant 
H\,{\sc ii}~regions in seven spiral galaxies NGC~628 (10), NGC~783 (8), 
NGC~2336 (28), NGC~6217 (3), NGC~6946 (39), NGC~7331 (4), and NGC~7678 (10). 
We also carried out $UBVRI$ photometry and H$\alpha$ spectrophotometry 
for the same SF~regions in the galaxies.

\begin{figure*}
\centerline{\includegraphics[width=15cm]{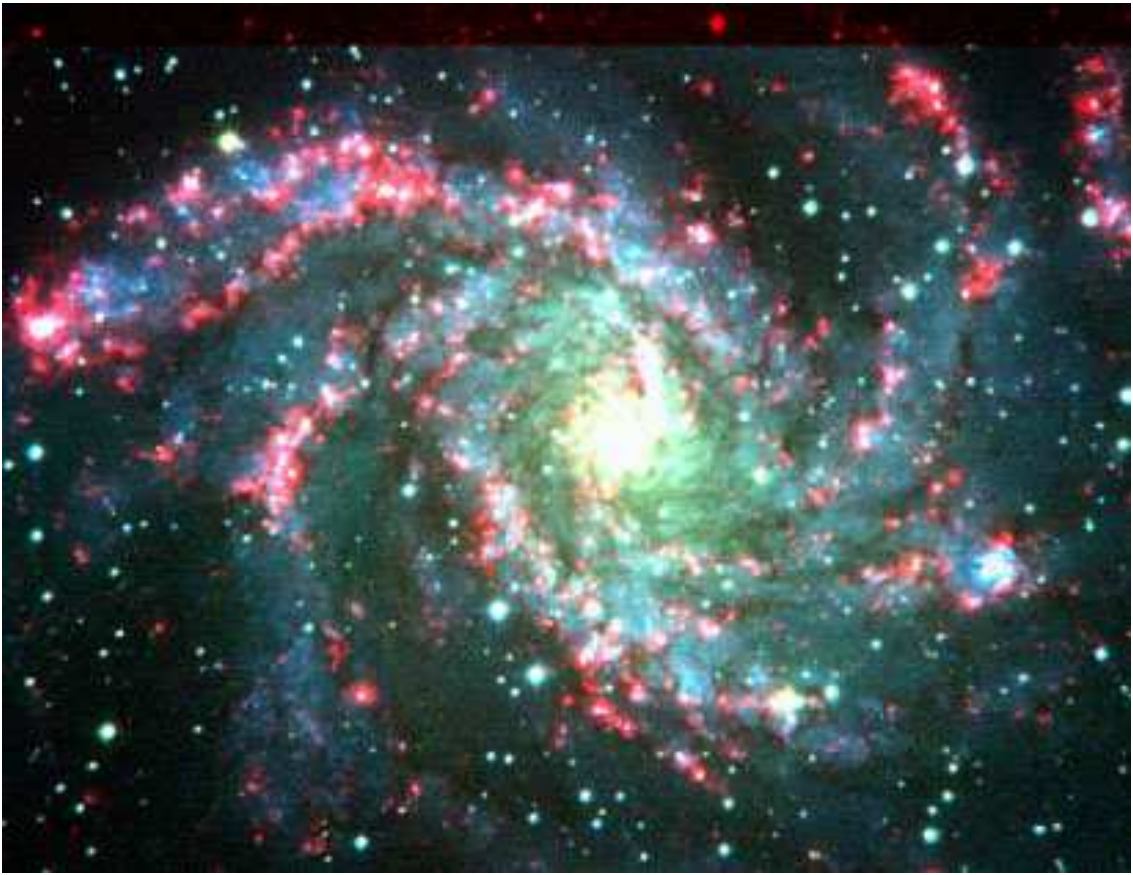}}
\caption{Gas and stars in NGC~6946. Red colour -- ionined gas (H$\alpha$ 
line image), blue colour -- young stars ($B$ band image), green colour -- 
older stars ($V$ band image). The size of the image is $8\times6$~arcmin$^2$ 
($13.6\times10.3$~kpc$^2$). 
\label{figure:n6946}}
\end{figure*}

Star forming region in an other galaxy is a single conglomerate of 
newly formed star clusters, dust clouds and ionized gas. The SF~region's 
sizes are within the range from several tens to $\sim1000$~pc; the ages 
are typically do not exceed 10~Myr \citep{elmegreen1996,efremov1998}. 
Being bright, SF~regions can be observed in nearby galaxies as objects 
of $16-20$ magnitudes in $U$ and $B$ bands with emission spectra, but 
they cannot be resolved into separate stars (see Fig.~\ref{figure:n6946}).

Accounting for the effects of gas and dust on observations of SF~regions 
is very important for the interpretation of multicolour photometry in 
terms of the initial mass function (IMF) and the star formation rate (SFR) 
history. Despite a huge amount of both spectroscopic and photometric 
observations of the extragalactic giant H\,{\sc ii}~regions -- SF~regions, 
the overlaps of the spectroscopic observations of SF~regions with the 
photometric ones are very poor. Using both techniques, we can eliminate the 
degeneracy between age and extinction, age and metallicity. Emission 
lines detected via spectroscopy can be used for disentangling the 
effects of extinction, accounting for the impact of the nebula emission 
on integrated broadband photometry and serve as valuable diagnostics of gas 
abundances \citep[see][and references therein]{sakhibov1990,reines2010}.

The galaxy sample is presented in Table~\ref{table:sample}, 
where fundamental parameters from the 
LEDA\footnote{http://leda.univ-lyon1.fr/} data base \citep{paturel2003} 
are provided.  The morphological type of galaxies  is given in column (2), 
the apparent and absolute magnitudes are listed in columns (3) and (4), 
the inclination and position angles -- in columns (5) and (6), and the 
isophotal radius in arcmin and kpc -- in columns (7) and (8). Adopted 
distances are given in the column (9). Finally, the Galactic absorption and 
the dust absorption due to the inclination of a galaxy are presented 
in columns (10) and (11). The Galactic absorptions, $A(B)_{\rm Gal}$, 
are taken from the NED\footnote{http://ned.ipac.caltech.edu/} data base. 
Other parameters are taken from the LEDA data base. Adopted value of the 
Hubble constant is equal to $H_0 = 75$ km\,s$^{-1}$Mpc$^{-1}$.

\begin{table*}
\caption[]{\label{table:sample}
The galaxy sample.
}
\begin{center}
\scriptsize{
\begin{tabular}{|c|c|c|c|c|c|c|c|c|c|c|c|} \hline \hline
NGC & Type & $B_t$ & $M_B^a$ & $i$ & PA & $R_{25}^b$ &
$R_{25}^b$ & $D$   & $A(B)_{\rm Gal}$ & $A(B)_{\rm in}$ & Ref.$^c$ \\
       &      & (mag) & (mag)   & (degr) & (degree) & (arcmin) &
(kpc)      & (Mpc) & (mag)        & (mag)       &           \\
1 & 2 & 3 & 4 & 5 & 6 & 7 & 8 & 9 & 10 & 11 & 12 \\
\hline
628   & Sc & 9.70 & --20.72 & 7 & 25 & 5.23 & 10.96 & 7.2 & 0.254 & 0.04 & 1 \\
783   & Sc & 13.18 & --22.01 & 43 & 57 & 0.71 & 14.56 & 70.5 & 0.222 & 0.45 & 2 \\
2336  & SB(R)bc & 11.19 & --22.14 & 55 & 175 & 2.51 & 23.51 & 32.2 & 0.120 & 0.41 & 3 \\
6217  & SB(R)bc & 11.89 & --20.45 & 33 & 162 & 1.15 & 6.89 & 20.6 & 0.158 & 0.22 & 4 \\
6946  & SABc & 9.75 & --20.68 & 31 &  62 & 7.74 & 13.28 & 5.9 & 1.241 & 0.04 & 5 \\
7331  & Sbc & 10.20 & --21.68 & 75 & 169 & 4.89 & 20.06 & 14.1 & 0.331 & 0.61 & 5 \\
7678  & SBc & 12.50 & --21.55 & 44 &  21 & 1.04 & 14.46 & 47.8 & 0.178 & 0.23 & 6 \\
\hline
\end{tabular}}
\end{center}
\begin{flushleft}
$^a$ Absolute magnitude of a galaxy corrected for Galactic extinction and
inclination effect. \\
$^b$ Radius of a galaxy at the isophotal level 25 mag/arcsec$^2$ in the
$B$ band corrected for Galactic extinction and inclination effect. \\
$^c$ References: 1 -- \citet{bruevich2007} and \citet{gusev2013b}, 
2 -- \citet{gusev2006a,gusev2006b}, 3 -- \citet{gusev2003}, 
4 -- \citet{artamonov1999} and \citet{gusev2015}, 5 -- \citet{gusev2015}, 
6 -- \citet{artamonov1997} and \citet{gusev2015}. \\
\end{flushleft}
\end{table*}

\section{Observations and reduction}

The spectroscopic observations were carried out in 2006--2008 at the 6~m 
telescope of Special Astrophysical Obsevatory (SAO) of the Russian Academy of 
Sciences with spectral camera attached at the focal reducer SCORPIO 
($f/4 \to f/2.6$) in the multislit mode; the field was about 6 arcmin and 
the pixel size was 0.178 arcsec on a EEV~42-40 ($2048\times2048$ pixels) 
CCD detector. We used the grism VPHG550G with a dispersion of 2.1\AA/pixel 
and a spectral resolution of 10\AA, which provided spectral coverage from 
[O\,{\sc ii}]$\lambda$3727+$\lambda$3729 oxygen emission lines to 
[S\,{\sc ii}]$\lambda$6717+$\lambda$6731 sulphur emission lines. As a 
result, we obtained 114 spectra for 102 H\,{\sc ii}~regions in seven 
galaxies. Detailed description of the observations and data reduction see 
in \citet{gusev2012,gusev2013a}.

The photometric observations of the galaxies were obtained in 1988--2006 with 
the 1~m and 1.5~m telescopes of the Maidanak Observatory (Uzbekistan) and 
1.8~m telescope of the Bohyunsan Optical Astronomy Observatory (South Korea). 
We used a photometric system close to the standard 
Johnson--Cousins $UBVRI$ system. The seeing during observations was 
$\sim1-1.5${\hbox{$^{\prime\prime}$}}. Detailed description of the 
observations and data reduction see in references in the last column of 
Table~\ref{table:sample}.

Spectrophotometric H$\alpha$ observations of three galaxies (NGC~628, 
NGC~6946, and NGC~7331) from our sample were made in 2006 with the 1.5~m 
telescope of the Mt.~Maidanak Observatory with the SI-4000 CCD camera. 
The wide-band interference H$\alpha$ filter ($\lambda_{eff}$ = 
6569\AA, FWHM = 44\AA) was used for the observations. The filter parameters 
provides H$\alpha$+[N\,{\sc ii}] imaging for these nearby galaxies. 
Detailed description of H$\alpha$ observations and data reduction see in 
\citet{gusev2013b} and \citet{gusev2016}.

The reduction of the spectroscopic, photometric and spectrophotometric data 
was carried out using standard techniques, with the European Southern 
Observatory Munich Image Data Analysis 
System\footnote{http://www.eso.org/sci/software/esomidas/} ({\sc eso-midas}).

We identified the SF~regions in NGC~628 and NGC~6946 using the list of 
H\,{\sc ii} regions of \citet{belley1992}. Identification of SF~regions in 
other galaxies was made by eye.

We took the geometric mean of the star forming complex major and 
minor axes for the SF~region's characteristic diameter $d$: 
$d = \sqrt{d_{max} \times d_{min}}$. We measured $d_{max}$ and $d_{min}$ 
from the radial $V$ profiles at the half-maximum brightness level (FWHM) 
for regions having a starlike profile, or by the distance between 
the points of maximum flux gradient for regions having extended (diffuse) 
profiles. We adopted the seeing as the error of the size measurements, which 
definitely exceeds all other errors.

Measurements of the apparent total $B$ magnitude and colour indices 
$U-B$, $B-V$, $V-R$, and $V-I$ were made within a round 
aperture. To measure $B$ magnitudes, we used the aperture size equal to 
the sum of the $d_{max}$ and the seeing in the $B$ band; in the case of 
measurements of colour indices, the aperture size is equal to the $d_{max}$. 
Use of smaller aperture for measurement of colour indices provides the 
spectral energy distribution in the brightest central part of SF~regions, 
i.e. in the young massive star clusters embedded in giant 
H\,{\sc ii}~regions. The galactic background is taken into account through 
the subtraction of the average flux, coming from the local surrounding 
region around the same area, from the flux, measured inside the round 
aperture.

We obtained magnitudes and colour indices for 101 of 102 studied SF~regions. 
One object in NGC~7331 is out of the image in the $UBVRI$ bands.

\section{Methods}

Combination of spectroscopic and multicolour photometric observations of 
SF~regions provides true colours and metallicities of young stellar 
groupings. That is necessary to account for the nebular emission impact 
and to eliminate ''age--extinction'' and ''age--metallicity'' degenerations 
in the comparative analysis with theoretical evolutionary models of star 
clusters.

SF~regions, studied here, constitute a single conglomerate of clouds of 
interstellar dust, ionised hydrogen, and newly formed star clusters. Much 
of the light emitted by stars in SF region's clusters undergoes absorption 
inside rather than outside the region itself. Therefore, if a SF~region 
contains the significant amount of the interstellar dust, then the light 
from SF~region stars is strongly attenuated even if the host galaxy is 
seen face-on. We assume that the emission from stars embedded in the 
SF~region is absorbed in the same way as the emission in lines of 
ionised hydrogen surrounding the star clusters in the SF~region. In other 
words, the light extinction for stars is equal to the light extinction for 
the emission of ionised gas $A{\rm (gas)} = A{\rm (stars)}$.

However, in some cases the gas extinction does not correspond to the 
absorption $A{\rm (stars)}$ in the young stellar associations (complexes). 
The purpose of our study was to find the criteria for which the condition 
$A{\rm (gas)} = A{\rm (stars)}$ is performed in the SF~region.

In the first, we determine the relative contributions of stellar 
continuum, nebular continuum, and emission lines fluxes to the total 
observed flux in the broadband filters \citep[see][]{sakhibov1990}.

\begin{figure*}
\vspace{2mm}
\centerline{\includegraphics[width=14cm]{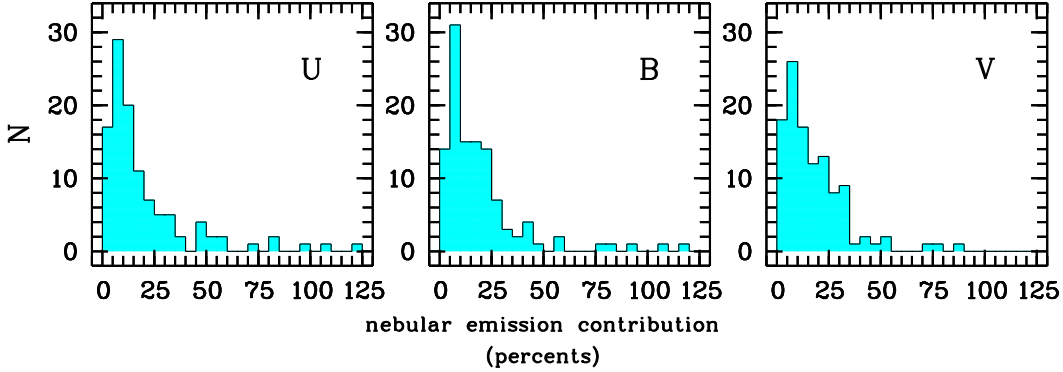}}
\caption{Distributions of studied SF~regions over nebular emission 
contribution in the $U$, $B$, and $V$ bands.
\label{figure:gas}}
\end{figure*}

Fig.~\ref{figure:gas} shows the distributions of studied SF~regions over 
relative contribution of the nebular emission to $U$, $B$, and $V$ fluxes. 
The number of objects decreases with the increase of relative 
contributions of the nebular emission. Some objects show extremely high 
nebular emission fluxes ($>40-50\%$), which are sometimes comparable or even 
more than total fluxes of SF~regions at least in one of $UBVR$ bands. The 
most part of these objects has extremely red colours (out of limits 
of model colours for the young stellar populations). However, according to 
\citet{reines2010} the objects with a nebular emission contribution more than 
$40-50\%$ in $B$, $V$ passbands have ages $<2$~Myr. We suggest that 
the condition $A({\rm stars})=A({\rm gas})$ is not valid in such SF~regions.

Some information about objects with very high nebular emission flux can be 
obtained from the comparison of the equivalent width EW(H$\alpha$), 
independently measured from our spectroscopic data, and the relation of the 
H$\alpha$ flux to star emission flux in the $R$ band, estimated from 
spectrophotometric and photometric observations.

\begin{figure*}
\vspace{2mm}
\centerline{\includegraphics[width=12cm]{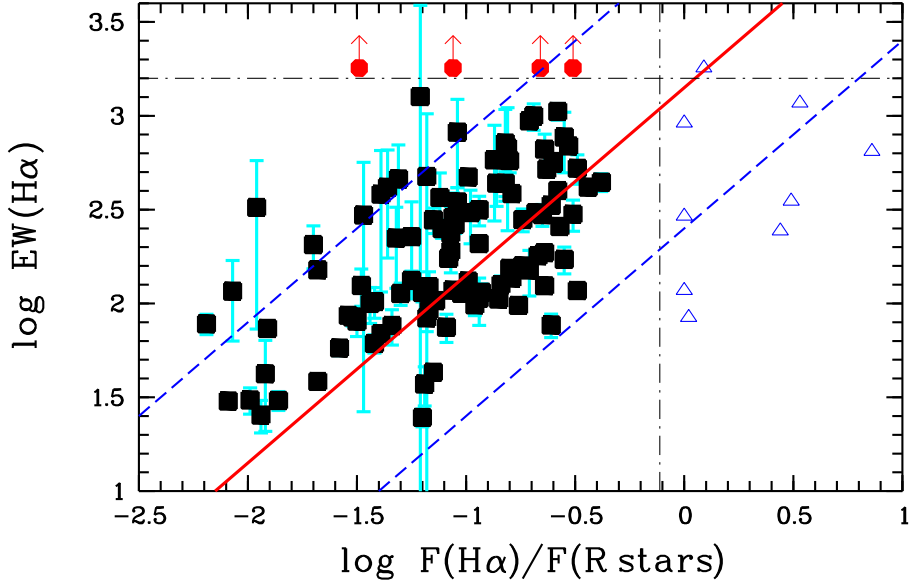}}
\caption{Diagram $\log F({\rm H}\alpha)/F(R_{\rm stars})$ versus 
$\log$~EW(H$\alpha$) for the SF~regions. Dot dashed lines are upper limits 
of the equivalent width EW(H$\alpha$) (horizontal line) and of the ratio 
of the H$\alpha$ flux to star emission flux in the $R$ band (vertical line), 
computed in \citet{reines2010}. The red solid line is a linear fit, 
computed for SF~regions (black squares), located under the upper limits of 
the EW(H$\alpha$) and $F({\rm H}\alpha)/F(R_{\rm stars}$) (horizontal and 
vertical dot dashed lines). Blue dashed lines are upper and lower $95\%$ 
prediction limits of the linear fit. Red circles show the SF~regions with an 
unreasonable ${\rm EW(H}\alpha)>1500$\AA. SF~regions with an unreasonable 
ratio $F({\rm H}\alpha)/F(R_{\rm stars})>1$ marked as blue triangles.
\label{figure:ewha}}
\end{figure*}

Fig.~\ref{figure:ewha} shows that most objects (black squares) located in 
the left lower part of the diagram under the upper limits of the equivalent 
width EW(H$\alpha$) (horizontal dot dashed line) and of the ratio of the 
H$\alpha$ flux to star emission flux in the $R$ band (vertical dot dashed 
line), computed in \citet{reines2010}. A distribution on the diagram of these 
SF~regions (black squares) can be described by the following linear 
regression (red solid line):
$\log{\rm EW}({\rm H}\alpha)=
\log(F({\rm H}\alpha)/F(R_{\rm stars}))+(3.15\pm0.05)$. Errors of values of 
EW(H$\alpha$) are used as weights by the linear fitting. The value of 
constant $3.15\pm0.05$ is in a good agremeent with the effective bandwidth 
($1580$\AA) and photometric zero-point of the $R$ filter determined by 
\citet{bessell1990} ($\approx3.1$).

Most part of objects which are located outside of this area (i.e. SF~regions 
with EW$({\rm H}\alpha)>1500$\AA\, and regions with an unreasonable ratio 
$F({\rm H}\alpha)/F(R_{\rm stars})>1$) shows extremely blue or extremely 
red $B-V$ colours in the colour-magnitude diagram (see below). It 
indicates either overestimated or underestimated light absorption for 
these objects. Note, that the discrepancy in estimations of the gas 
contribution in the SF~regions obtained from the $R$ and 
H$\alpha$ photometry and spectroscopic data may indicate the spatial 
deviation between positions of the H\,{\sc ii}~region and the star cluster 
associated with it. For these SF~regions the slit of spectrograph crosses 
the centre of H\,{\sc ii} region, but covers the edge of star cluster. 
\citet{maizapellaniz1998} obtained the two-dimensional spectrophotometric 
map of the central region of NGC~4214 and showed that stars, gas, and dust 
clouds in the brightest SF~regions near the galactic nucleus are spatially 
separated. The dust is concentrated at the edges of the region of ionization 
and primarily influences nebular emission lines, whereas the stellar 
continuum is located in a region that is relatively free of dust and gas. 
Thus, the adopted here, assumption of $A({\rm stars})=A({\rm gas})$ is not 
valid in such SF~regions.

\section{Results}

We estimated the oxygen and nitrogen abundances and the electron 
temperatures in 80 out of 102 H\,{\sc ii}~regions through the recent variant 
of the strong-line method \citep[NS calibration;][]{pilyugin2011}. 

The parameters of the radial distribution (the extrapolated central intercept 
value and the gradient) of the oxygen and nitrogen abundances in the discs of 
six galaxies (all except NGC~6217) have been determined. The abundances in 
NGC~783, NGC~2336, NGC~6217, and NGC~7678 are measured for the first time. 
The extrapolated central oxygen and nitrogen abundances in studied galaxies 
are in the range from 8.73 to 8.94, and from 8.21 to 8.49, respectively 
(see Table~\ref{table:abund}); radial gradients of oxygen and nitrogen 
abundances are in the range from --0.65 to --0.38 dex/$R_{25}$, and from 
--1.34 to --0.90 dex/$R_{25}$, respectively. The only exception is SBc 
galaxy NGC~7678 with flat radial distribution of oxygen and nitrogen
(Table~\ref{table:abund}).

\begin{table*}
\caption[]{\label{table:abund}
Parameters of radial distributions of oxygen and nitrogen abundances 
in the discs of galaxies.
}
\begin{center}
\begin{tabular}{|c|c|c|c|c|} \hline \hline
Galaxy & O/H & O/H & N/H & N/H \\
 & center$^a$ & gradient$^b$ & center$^a$ & gradient$^b$ \\ 
\hline
NGC  628 & 8.74$\pm$0.02 & --0.43$\pm$0.03 & 8.26$\pm$0.05 & --1.34$\pm$0.08 \\
NGC  783 & 8.94$\pm$0.12 & --0.65$\pm$0.16 & 8.49$\pm$0.19 & --1.07$\pm$0.26 \\
NGC 2336 & 8.79$\pm$0.05 & --0.38$\pm$0.07 & 8.25$\pm$0.10 & --0.90$\pm$0.15 \\
NGC 6946 & 8.73$\pm$0.02 & --0.43$\pm$0.06 & 8.21$\pm$0.08 & --1.25$\pm$0.19 \\
NGC 7331 & 8.75$\pm$0.18 & --0.49$\pm$0.36 & 8.23$\pm$0.36 & --1.03$\pm$0.74 \\
NGC 7678 & 8.61$\pm$0.03 & --0.15$\pm$0.05 & 7.89$\pm$0.17 & --0.34$\pm$0.27 \\
\hline
\end{tabular}
\end{center}
\begin{flushleft}
$^a$ in unit of 12+log(X/H). $^b$ in unit of dex/$R_{25}$. \\
\end{flushleft}
\end{table*}

\begin{figure*}
\vspace{2mm}
\centerline{\includegraphics[width=10cm]{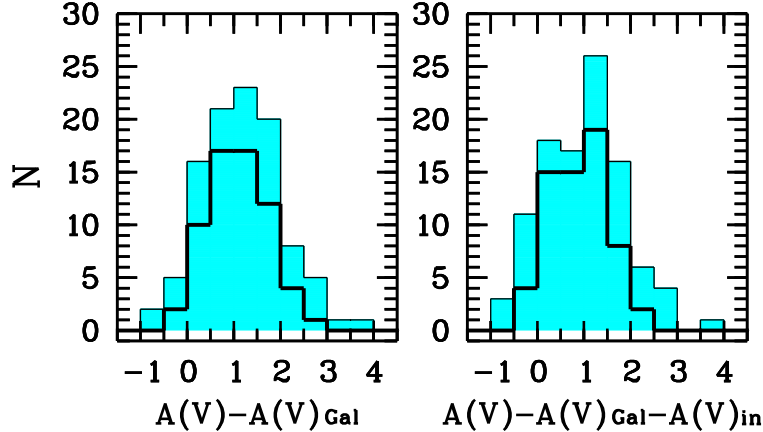}}
\caption{Frequency distribution of extinction in H\,{\sc ii}~regions, 
$A(V)$ corrected for the Galactic absorption $A(V)_{\rm Gal}$ (left), 
and for both the Galactic absorption and the dust absorption due to the 
inclination of a galaxy $A(V)_{\rm Gal}+A(V)_{\rm in}$ (right). The cyan 
histogram is the distribution of sample of 102 SF~regions. The black 
histogram is the distribution of sample of 63 SF~regions in which the 
condition $A({\rm stars})=A({\rm gas})$ is satisfied.
\label{figure:av}}
\end{figure*}

All galaxies from our sample follow well the general trend in the 
''luminosity -- central metallicity'' diagram for spiral and irregular 
galaxies.

Extinction in studied H\,{\sc ii}~regions, obtained from the Balmer 
decrement, varies greatly, from 0 up to 3~mag. Frequency distributions of 
extinction $A(V)$ in SF~regions corrected for the Galactic absorption 
$A(V)_{\rm Gal}$ and the dust absorption due to the inclination of a galaxy 
$A(V)_{\rm in}$ are presented in Fig.~\ref{figure:av}. The 
distributions show Gauss profiles with the mean 
$A(V)-A(V)_{\rm Gal} = 1.18\pm0.84$~mag and 
$A(V)-A(V)_{\rm Gal}-A(V)_{\rm in} = 1.01\pm0.86$~mag. This result is in a 
good agreement with reddening measurements for the sample of 49 disc, halo 
and nuclear star clusters in M82 \citep{konstantopoulos2009}.

For the further study we excluded SF~regions in which 
$A({\rm stars}) \ne A({\rm gas})$, or $A(V)$ was overestimated or 
underestimated (i.e. SF~regions with the contribution to the total emission 
of gas in the $B$ and/or $V$ passbands is higher than $40\%$, 
EW$({\rm H}\alpha)>1500$\AA, unreasonable ratio 
$F({\rm H}\alpha)/F(R_{\rm stars})>1$, and SF~regions with the spatial 
deviation between positions of the H\,{\sc ii}~region and the star cluster 
associated with it). We also excluded SF~regions with large errors of 
$A(V)$ estimations. We believe that the condition 
$A({\rm stars})=A({\rm gas})$ is satisfied in the remaining 63 SF~regions.

The mean extinction for the sample of 63 SF~regions in which 
$A({\rm stars})=A({\rm gas})$ is $A(V)-A(V)_{\rm Gal} = 1.09\pm0.63$~mag and 
$A(V)-A(V)_{\rm Gal}-A(V)_{\rm in} = 0.91\pm0.62$~mag.

Two objects in NGC~6946 have unreasonable value 
$A(V)+\Delta A(V)<A(V)_{\rm Gal}$ (Fig.~\ref{figure:av}). Apparently, we 
underestimated the values of $A(V)$ in these regions. Remark that one of 
them has the nebular emission contribution in the $B$ band $>40\%$, and the 
second region has EW$({\rm H}\alpha)>1500$\AA.

\begin{figure*}
\vspace{4mm}
\centerline{\includegraphics[width=14cm]{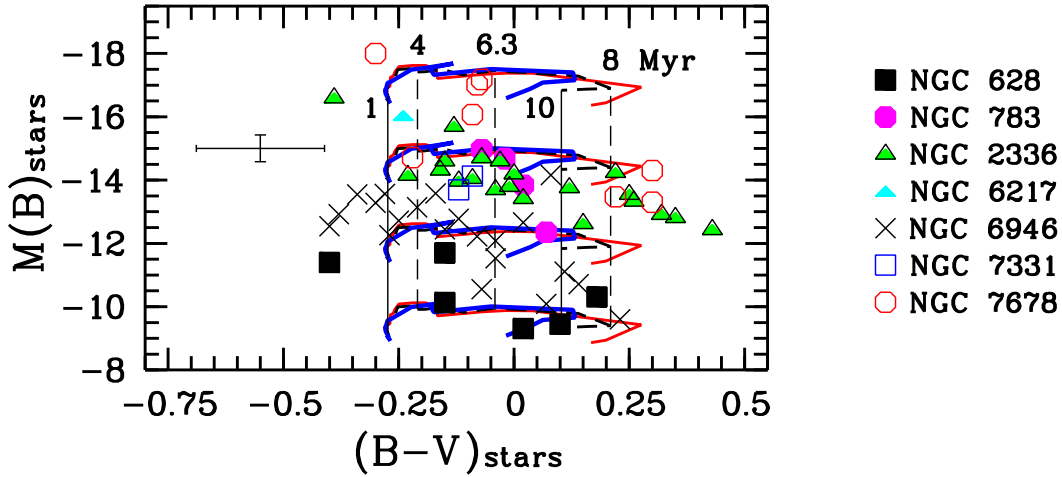}}
\caption{True colours and luminosities of 63 SF~regions compared with 
SSP models in the colour-magnitude diagram. Models with $Z=0.008$ (blue), 
0.012 (black dashed), and 0.019 (red), with ages from 1 to 10~Myr, and 
with masses $10^4 M_{\odot}$, $10^5 M_{\odot}$, $10^6 M_{\odot}$, and 
$10^7 M_{\odot}$ are shown. Isochrones with ages 1, 10 (solid horizontal
lines), 4, 6.3, and 8~Myr (dashed horizontal lines) for the evolutionary 
tracks with $Z=0.012$ are shown. The error's cross shows the mean accuracy 
of $M(B)_{\rm stars}$ and $(B-V)_{\rm stars}$ of these objects. See the 
text for details.
\label{figure:cmd}}
\end{figure*}

In Fig.~\ref{figure:cmd} we compare the true colours and luminosities of 
63 SF~regions in galaxies with SSP models with Salpeter IMF slope 
$\alpha=−2.35$ in the colour-magnitude diagram. The age of models, shown 
in the diagram, ranges from 1~Myr up to 10~Myr. The figure shows that the 
objects occupate the models area. Displacements of the objects from the 
models area lie within the errors interval. 

Several objects in distant galaxies NGC~2336 and NGC~7678 are located out 
of area of young models ($t<10$~Myr), and can be fitted with older models. 
As sizes of all these objects belong to the range from 300~pc till 1000~pc, 
they are cluster complexes. There are two possible interpretation of their 
location. Firstly, it may indicate the underestimated light extinction and 
$A({\rm stars})>A({\rm gas})$. Secondly, it may indicate on possible 
coexistence of old and extremely young star clusters in the same SF~region.

We estimated masses and ages for 63 SF~regions. The regions have 
ages from 1 to 10~Myr, the most part of them are younger than 6~Myr 
(Fig.~\ref{figure:cmd}). The derived masses of young stellar 
groupings range from $10^4 M_{\odot}$ in the nearby galaxies NGC~628 and 
NGC~6946 to $10^7 M_{\odot}$ in the most distant NGC~7678. More than 
$80\%$ of groupings have masses between $10^5 M_{\odot}$ and 
$10^6 M_{\odot}$ (see Fig.~\ref{figure:cmd}). The lowest mass estimate of 
$\approx1\cdot10^4 M_{\odot}$ for the objects in NGC~628 and NGC~6946 
belongs to the mass interval of the youngest Galactic open clusters.

\section{Conclusions}

We have presented a combination of the spectroscopic and photometric studies 
of the disc cluster population in seven spiral galaxies. Our primary goal 
was to derive spectroscopic information on star forming regions (extinction, 
chemical abundance, relative contributions of nebular continuum and 
emission lines to the total observed flux) and photometric information on 
young stellar groupings (true colours, luminosity, mass, age) embedded in 
H\,{\sc ii}~regions.

We estimated the oxygen and nitrogen abundances and the electron 
temperatures in 80 out of 102 H\,{\sc ii}~regions in seven spiral galaxies.

We determined the parameters of the radial distribution of the oxygen and 
nitrogen abundances in the discs of six galaxies. The abundances in 
NGC~783, NGC~2336, NGC~6217, and NGC~7678 are measured for the first time.

We obtained $UBVRI$ magnitudes and colour indices for 101 SF~regions.

We found the criteria for which the condition 
$A({\rm stars})=A({\rm gas})$ is not performed in the SF~region.

We estimated masses and ages for 63 star forming regions. The 
regions have ages from 1 to 10~Myr, the most part of them are younger 
than 6~Myr. The derived masses of young stellar groupings range from 
$10^4 M_{\odot}$ to $10^7 M_{\odot}$. More than $80\%$ of groupings have 
masses between $10^5 M_{\odot}$ and $10^6 M_{\odot}$.

More detailed description of this work see in 
\citet{gusev2012,gusev2013a,gusev2016}.

\section*{Acknowledgement}

The authors acknowledge the usage of the HyperLeda data base 
(http://leda.univ-lyon1.fr), the NASA/IPAC Extragalactic Database 
(http://ned.ipac.caltech.edu), and the Padova group online server CMD 
(http://stev.oapd.inaf.it). This study was supported by the Russian 
Science Foundation (project no. 14--22--00041).


\begin{thebibliography}{}

\bibitem[\protect\citeauthoryear{Artamonov et al.}{1997}]{artamonov1997}
          Artamonov B.~P., Bruevich V.~V., Gusev A.~S. (1997) ARep 41, 577

\bibitem[\protect\citeauthoryear{Artamonov et al.}{1999}]{artamonov1999}
          Artamonov B.~P., Badan Yu.~Yu., Bruyevich V.~V., 
          Gusev~A.S. (1999) ARep 43, 377

\bibitem [\protect\citeauthoryear{Belley and Roy}{1992}]{belley1992}
          Belley J., Roy J.-R. (1992) ApJS 78, 61

\bibitem [\protect\citeauthoryear{Bessell}{1990}]{bessell1990}
          Bessell M.~S. (1990) PASP 102, 1181

\bibitem [\protect\citeauthoryear{Bruevich et al.}{2007}]{bruevich2007}
          Bruevich V.~V., Gusev A.~S., Ezhkova O.~V. et al. (2007) 
          ARep 51, 222

\bibitem [\protect\citeauthoryear{Efremov and Elmegreen}{1998}]{efremov1998}
          Efremov Yu.~N., Elmegreen B.~G. (1998) MNRAS 299, 588

\bibitem [\protect\citeauthoryear{Elmegreen and Efremov}{1996}]{elmegreen1996}
          Elmegreen B.~G. Efremov Yu.~N. (1996) ApJ 466, 802

\bibitem [\protect\citeauthoryear{Gusev}{2006a}]{gusev2006a}
          Gusev A.~S. (2006a) ARep 50, 167

\bibitem [\protect\citeauthoryear{Gusev}{2006b}]{gusev2006b}
          Gusev A.~S. (2006b) ARep 50, 182

\bibitem [\protect\citeauthoryear{Gusev and Efremov}{2013}]{gusev2013b}
          Gusev A.~S., Efremov Yu.~N. (2013) MNRAS 434, 313

\bibitem [\protect\citeauthoryear{Gusev and Park}{2003}]{gusev2003}
          Gusev A.~S., Park M.-G. (2003) A\&A 410, 117

\bibitem [\protect\citeauthoryear{Gusev et al.}{2012}]{gusev2012}
          Gusev A.~S., Pilyugin L.~S., Sakhibov F. et al. (2012) 
          MNRAS 424, 1930

\bibitem [\protect\citeauthoryear{Gusev et al.}{2013}]{gusev2013a}
          Gusev A.~S., Sakhibov F.~H., Dodonov S.~N. (2013)
          AstBull 68, 40

\bibitem [\protect\citeauthoryear{Gusev et al.}{2015}]{gusev2015}
          Gusev A.~S., Gyslyakova S.~A., Novikova A.~P. et al. (2015) 
          ARep 59, 899

\bibitem [\protect\citeauthoryear{Gusev et al.}{2016}]{gusev2016}
          Gusev A.~S., Sakhibov F., Piskunov A.~E. et al. (2016) 
          MNRAS submitted

\bibitem [\protect\citeauthoryear{Konstantopoulos et al.}{2009}]{konstantopoulos2009}
          Konstantopoulos I.~S., Bastian N., Smith L.~J. et al. 
          (2009) ApJ 701, 1015

\bibitem [\protect\citeauthoryear{Ma\'{i}z-Apell\'{a}niz et al.}{1998}]{maizapellaniz1998}
          Ma\'{i}z-Apell\'{a}niz J., Mas-Hesse J.~M., 
          Mu\~{n}oz-Tu\~{n}\'{o}n C. et al. (1998) A\&A 329, 409

\bibitem [\protect\citeauthoryear{Paturel at al.}{2003}]{paturel2003}
          Paturel G., Petit C., Prugniel Ph. et al. (2003) A\&A 412, 45

\bibitem [\protect\citeauthoryear{Reines et al.}{2010}]{reines2010}
          Reines A.~E., Nidever D.~L., Whelan D.~G., Johnson K.~E. 
          (2010) ApJ 708, 26

\bibitem [\protect\citeauthoryear{Pilyugin and Mattsson}{2011}]{pilyugin2011}
          Pilyugin L.~S., Mattsson L. (2011) MNRAS 412, 1145

\bibitem [\protect\citeauthoryear{Sakhibov and Smirnov}{1990}]{sakhibov1990}
          Sakhibov F., Smirnov M.~A. (1990) SvA 34, 236

\end{thebibliography}
\end{document}